\documentclass[reprint,twocolumn,secnumarabic,amssymb, nobibnotes, aps, prd, showpacs]{revtex4-1}
\usepackage{graphicx}% Include figure files
\usepackage{dcolumn}% Align table columns on decimal point
\usepackage{bm}% bold math
\usepackage{slashed}
\begin{document}

\preprint{AIP/123-QED}

\title[]{Low energy kaon-hyperon interaction}% Force line breaks with \\
\author{M. G. L. N. Santos}
 \email{magwwo@gmail.com}%Lines break automatically or can be forced with \\
\author{C. C. Barros, Jr}%
 \email{barros.celso@ufsc.br}
\affiliation{ 
Departamento de F{\'{i}}sica, CFM, Universidade Federal de Santa Catarina\\ Florian{\'{o}}polis SC, CEP 88010-900, Brazil%\\This line break forced with \textbackslash\textbackslash
}%
%%%%%%%%%%%%%%%%%%%%%%%%%%%%%%%%%%%%

\date{\today} % Leave empty to omit a date

\begin{abstract}
In this work we
study the low energy kaon-hyperon interaction considering effective chiral Lagrangians 
that include kaons, $\sigma$ mesons, hyperons and the corresponding 
resonances.
We calculate the scattering amplitudes, and then the
total cross sections, angular distributions, polarizations and the $S$ 
and $P$ phase shifts. 
\end{abstract}

 % mesmo do pion-hyperon
\maketitle
\section{Introduction}

Until today a subject that is very interesting and remains not very well studied is the 
low energy hyperon interactions. Despite the fact that experimental data for many
hyperon processes are not available (as for example the
$K\Lambda$ and $K\Omega$ interactions) 
and that by the theoretical side they are not fully described, this kind of interaction is 
a fundamental element for several physical systems of interest.

In the study of the hypernuclei structure \cite{hnuc1}-\cite{hnuc4}, the knowledge 
of the 
nucleon-hyperon and hyperon-hyperon
interactions is an essential aspect. In order to understand these interactions,
and to determine the potentials of interest, an accurate understanding of 
the meson-hyperon interactions is needed.  

Another system where hyperon interactions are required is in the study of the hyperon stars.
After the proposal of the hypothesis where hyperons could be produced inside  
neutron stars at high densities,
many models have been proposed, as for example in
\cite{hipstar1}-\cite{hipstar4}, and the effect of the
presence of hyperons in the equations of state, and consequently in the
determination of the star masses
have been studied. The indeterminations
in the nucleon-hyperon and hyperon-hyperon potentials cause difficulties in the
understanding of these stars.

In high energy physics this kind of
interaction is very important also. When studying
hyperon polarization, produced in proton-nucleus and nucleus-nucleus collisions 
\cite{RH1}-\cite{mor}, 
in \cite{cy}-\cite{ccb2} the final interactions of the hyperons
and antihyperons
with the produced pions is a central ingredient in order to explain the final 
polarizations. 
As it has been shown, the effect of the hyperon interactions
with the surrounding hot medium, composed predominantly of pions, is
very important. 
The observed differences between the
polarizations of hyperons and antihyperons are
very difficult to be explained in another way.
The effect of the final kaon-hyperon interactions
has not been considered yet, and it may cause corrections in the
final polarization.
For 
this reason, this work is very important and this effect must be
investigated. 
Recent results from RHIC \cite{plhc1} and even the hyperons
produced in the LHC may 
 be studied in a similar form, and in order to obtain accurate results
these interactions must be considered.

For these reasons, this work will be devoted to the study of the kaon-hyperon ($KY$) 
and antikaon-hyperon (${\overline K}Y$) interactions.
This work may be considered as a continuation of the study proposed in \cite{BH}-\cite{ccb}, where
the pion-hyperon interactions have been described with a model based in effective 
chiral
Lagrangians where 
the exchange of mesons and baryons has been taken into account. In this model \cite{BH}, 
the resonances
dominate many channels of the reactions as it may be seen in the results.
This behavior may be considered as an experimental feature,
 fact that is similar to what happens in 
the low energy pion-nucleon interactions, where the isospin 3/2 and spin 3/2 channel is 
dominated by the $\Delta^{++}$ resonance. Comparison with the data from the 
HyperCP experiment
\cite{hyper1}, \cite{hyper2}
shows a very good accord with the results obtained for the
 $\pi\Lambda$ scattering in \cite{BH}.
 So, the work that will be shown in this paper
  is based on the ideas presented in this model.

This paper will present the following content: in Sec. II, the basic formalism will be shown, 
in Sec. III, the kaon-lambda ($K\Lambda$) interactions will be studied, in Sec. IV, the antikaon-lambda
(${\overline K}\Lambda$), and in Sec. V, the kaon-sigma will be shown. In Sec. VI we will
present the antikaon-sigma interactions (${\overline K}\Sigma$) and finally, the discussions and 
conclusions in Sec. VII. In the Appendix, some expressions of interest will be presented.

%%%%%%%%%%%%%%%%%%%%%%%%%%%%%%%%%%%%%%%%%%%%%%%%%%%%%%%%%%%%%%%%%%%%%%%%%%%%%%%%%%%%%%%%%%%%%%%%%%%%%%%%%%%%%%%%%%%%%%%%%%%%%%%%%%%%
\section{The Method}

In order to study the $KY$ and ${\overline K}Y$ interactions, we will use a model
 proposed with the purpose of
studying the low energy pion-hyperon interactions  \cite{BH}-\cite{ccb}, that is based in an
analogy with models successfully used to describe
the $\pi N$ interactions considering chiral effective Lagrangians.
These interactions are very well studied,
as for example in \cite{manc}-\cite{pi2}, 
both theoretically, where many models have been proposed, and experimentally, with 
a large amount of experimental data available. A basic characteristic of this 
system is the dominance of resonances in the scattering amplitudes at low energies.
The $\Delta^{++}$, for example, dominates the cross section of the $\pi^+p$ scattering
at low energies. As this particle has spin 3/2 and isospin 3/2, it may be introduced in the theory 
by considering a Lagrangian in the form of eq.~(\ref{eq2}). In the study of the 
pion-hyperon 
interactions \cite{BH}, a similar behavior has been observed, so we expect that in 
the $KY$ interactions it also occurs.

In this section we will present the basic formalism that will be used to
study the kaon-hyperon interactions (that is the same one 
worked out in the study the 
pion-hyperon interactions \cite{BH}) and how the observables may be obtained.
In the method that will be followed
 in this work, some important characteristics of the interacting particles
will be
implemented, the spin, the isospin, and the masses of each one of them. 
These characteristics determine which Lagrangian have to be used in order to build the model.

For example, in \cite{manc}, the
 Lagrangians considered 
to study the $\pi N$ scattering are given by 
\begin{equation}
\label{eq1}
\mathcal{L}_{\pi NN}= \frac{g}{2m}\big(\overline{N}\gamma_\mu\gamma_5\vec{\tau}N\big)\cdot\partial^\mu\vec{\phi}\ ,
\end{equation}

\begin{equation}
\label{eq2}
\mathcal{L}_{\pi N\Delta}=  g_\Delta\Big\{\overline{\Delta}^\mu\big[g_{\mu\nu}-(Z+\frac{1}{2})\gamma_\mu\gamma_\nu\big]\vec{M}N\Big\}\cdot\partial^\nu\vec{\phi}\ ,
\end{equation}

\begin{eqnarray}
\mathcal{L}_{\rho NN}&=& \frac{ g_0}{2}\Big[\overline{N}\gamma_\mu\vec{\tau}N\Big]\cdot\vec{\rho^\mu}
\label{eq3}
+\frac{ g_0}{2}\Big[\overline{N}\Big(\frac{\mu_p-\mu_n}{4m}\Big)i\sigma_{\mu\nu}\vec{\tau}N\Big]\nonumber\\
&&\times\big(\partial^\mu\vec{\rho^\nu}-\partial^\nu\vec{\rho^\mu}\big)\ ,
\end{eqnarray}

\begin{equation}
\mathcal{L}_{\pi\rho\pi}=g_0\vec{\rho^\mu}\cdot\Big(\vec{\phi}\times\partial_\mu\vec{\phi}\Big)
\label{eq4}
-\frac{g_0}{4m^2_\rho}\big(\partial_\mu\vec{\rho_\nu}-\partial_\nu\vec{\rho_\mu}\big)\cdot\Big(\partial^\mu\vec{\phi}\times\partial^\nu\vec{\phi}\Big)\ ,
\end{equation}
where $N$, $\Delta$, $\vec{\phi}$, $\vec{\rho}$ are the nucleon, delta, pion, and rho fields with masses $m$, $m_\Delta$, $m_\pi$, and $m_\rho$, respectively, $\mu_P$ and $\mu_n$ are the proton and neutron magnetic moments \cite{pdg}, $\vec{M}$ and $\vec{\tau}$ are the isospin recombination
matrices, and $Z$ is a parameter representing the possibility of the off-shell-$\Delta$ having spin 1/2. The parameters $g$, $g_\Delta$ and $g_0$ are 
the coupling constants. In \cite{BH} similar Lagrangians have been
used to study the pion-hyperon interactions, and in this work the
same procedure will be adopted. So, in the following sections these Lagrangians 
will be adapted to the kaon-hyperon systems.

Calculating the diagrams, considering the interactions described by the 
Lagrangians above for an arbitrary process,
 the scattering amplitudes  may be written in the form
\begin{equation}
T^{\beta\alpha}_{\pi N}=\sum_IT^I\left\langle\beta|P_I|\alpha\right\rangle
=\sum_IT^IP_I^{\beta\alpha} \ ,
\label{eq5}
\end{equation}
\noindent
that is a sum over all the $I$ isospin states where
$P_I$ is a projection operator,the indices $\alpha$ and $\beta$ are relative to
the initial and final isospin states of the $\pi$, and $T^I$ is an amplitude for a given isospin state
that may be written as
\begin{equation}
T^I=\overline{u}(\vec{p'})\Big[A^{I}+\frac{1}{2}(\slashed{k}+\slashed{k}')B^I\Big]u(\vec{p})
\   ,
\label{eq:}
\end{equation}
where
 $u(\vec{p})$ is a spinor representing the initial baryon, incoming with 
four-momentum $p_\mu$. The final baryon has a spinor $\overline u(\vec{p'})$, four-momentum $p_\mu'$,
and $k_\mu$ and $k_\mu '$ are the incoming and outgoing
meson four-momenta.
The amplitudes $A^I$ and $B^I$ are calculated from the diagrams.
So, if these amplitudes are determined, the $T^I$ amplitudes 
may be obtained and then we will be
able to compute the observables of interest.  
 
The scattering matrix for an isospin state is
given by the expression
\begin{equation}
M^I=\frac{T^I}{8\pi \sqrt{s}} \ ,
\end{equation}
which may be decomposed into the spin-non-flip and spin-flip amplitudes $f^I(k,\theta)$ and $g^I(k,\theta)$, defined in terms of the momentum
$k=|\vec k|$ and $x=\cos\theta$, $\theta$
the scattering angle, as
\begin{equation}
M^I=f^I(k,x)+ g^I(k,x) \vec{\sigma}.\hat{n}\ ,
\end{equation}

\noindent
where $\hat n$ is a vector normal to the scattering plane,
and may be  expanded in terms of the partial-wave amplitudes $a_{l\pm}$ with
\begin{equation}
f^I(k,x)=\sum_{l=0}^\infty{\Big[(l+1)a_{l+}^I(k)+la_{l-}^I(k)\Big] P_l (x)}\ ,
\end{equation}

\begin{equation}
g^I(k,x)=i\sum_{l=1}^\infty{\Big[a_{l-}^I(k)-a_{l+}^I(k)\Big] P_l^{(1)} (x)}\  .
\end{equation}
These amplitudes may be calculated using the Legendre polynomials
 orthogonality relations
\begin{equation}
a_{l\pm}^I(k)=
\frac{1}{2}\int_{-1}^1\Big[P_l(x)f^I_1(k,x) +P_{l\pm 1}(x)f^I_2(k,x)\Big] dx \ ,
\end{equation}
with	
\begin{equation}
f_1^{I}(k,x)=\frac{(E+m)}{8\pi \sqrt{s}}[A^{I}+(\sqrt{s}-m)B^I]\ ,
\end{equation}
\begin{equation}
f_2^{I}(k,x)=\frac{(E-m)}{8\pi \sqrt{s}}[-A^{I}+(\sqrt{s}+m)B^I]\ ,
\end{equation}
where $E$ is the baryon energy in the center-of-mass frame
and $\sqrt{s}$ is given by a Mandelstam variable
 (see the Appendix). At low energies the  $S$ ($l=$0) and $P$ ($l=$1) waves 
dominate the scattering amplitudes, and for higher values of $l$ 
the amplitudes  are 
much smaller (almost negligible), so they
 may be considered as small corrections.

Calculating the amplitudes at the tree-level, the results obtained will be real, 
and then violate the unitarity of the $S$ matrix. 
As it is usually done, we may reinterpret these results
as elements of the $K$ reaction matrix \cite{BH}-\cite{ccb}
and then obtain unitarized amplitudes    
\begin{equation}
a_{l\pm}^U=\frac{a_{l\pm}}{1-ik  a_{l\pm}}   \ .
\end{equation}

The differential cross sections may be calculated
using the previous results
\begin{equation}
\frac{d\sigma}{d\Omega}=|f|^2+|g|^2
  \  ,
\label{eq:}
\end{equation}

\noindent
and integrating this expression
over the solid angle we obtain the total cross sections 
\begin{equation}
\sigma_T=4\pi \sum_l{\Big[(l+1)|a_{l+}^U|^2+l|a_{l-}^U|^2\Big]}   \  .
\end{equation}

\noindent
The phase shifts are given by
\begin{equation}
\delta_{l\pm}=\tan^{-1}(ka_\pm)
  \ ,
\label{eq:}
\end{equation}

\noindent
and finally, the polarization , 
\begin{equation}
\vec{P}=-2 \frac{Im(f^*g)}{|f|^2+|g|^2}\hat{n}
  \  .
\label{eq:}
\end{equation}

\noindent
An important task to achieve is to determine the coupling constant for each resonance 
that will be considered. We will adopt the same procedure considered in
 \cite{BH}, comparing the
amplitude obtained in the calculations with the relativistic Breit-Wigner expression,
that is determined in terms of experimental quantities
\begin{equation}
\delta_{l\pm}=\tan^{-1}\Bigg[\frac{\Gamma_0\Big(\frac{k}{k_0}\Big)^{2J+1}}{2(m_r-\sqrt{s})}\Bigg]\ ,
\label{eq19}
\end{equation}
where $\Gamma_0$ is the width, $k_0=|\vec{k_0}|$ is the  momentum at the peak of the 
resonance in the
center-of-mass system,
$m_r$ is its mass and $J$ the angular momentum,
 considering the data from \cite{pdg}. We will consider the coupling constant
that better fits this experession in each case.

In the following sections we will apply this formalism in the study of the reactions
 of interest.

%%%%%%%%%%%%%%%%%%%%%%%%%%%%%%%%%%%%%%%%%%%%%%%%%%%%%%%%%%%%%%%%%%%%%%%%%%%%%%%%%%%%%%%%%%%%%%%%%%%%%%%%%%%%%%%%%%%%%%%%%%%%%%%%%%%

\section{Kaon-Lambda Interaction}

Since the $\Lambda$ hyperon has isospin 0, the scattering amplitude for the 
 $K\Lambda$ interaction will have the form
\begin{equation}
T_{K\Lambda}=\bar{u}(\vec{p'})\Big[A(k,\theta)+\Big(\frac{\slashed k+\slashed k'}{2}\Big)B(k,\theta)\Big]u(\vec{p})
   \  ,
\label{eq20}
\end{equation}
with the variables defined in  section II. 
Comparing this expression with (\ref{eq5}), we
have a simple result,
$P_{1/2}^{\beta\alpha}=1$, as the kaon has isospin $1/2$, and just
one isospin amplitude.

\begin{figure}[!htb]
\includegraphics[width=0.50\textwidth]{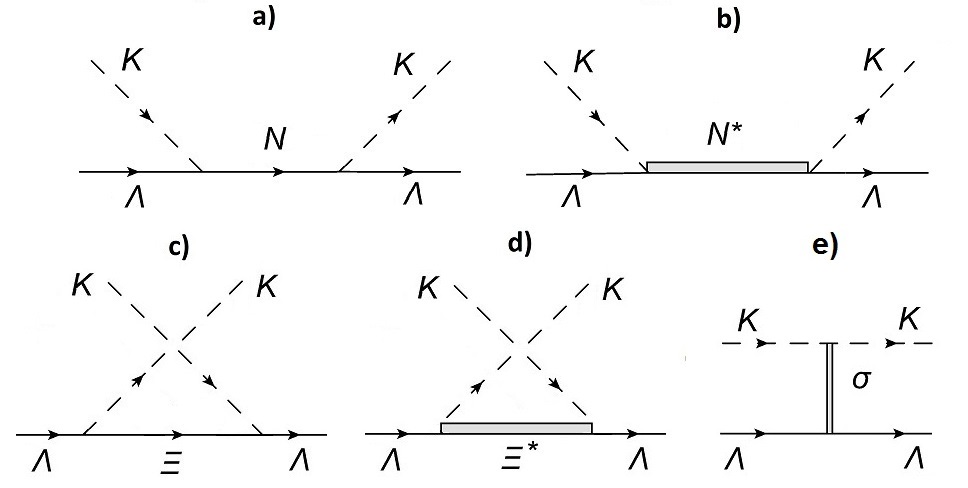}
\caption{Diagrams for the $K\Lambda$ interation}\label{fig1}
\end{figure}
 
In FIG.~\ref{fig1} we show the diagrams and the particles considered 
to formulate
 the $K\Lambda$ interaction. The particles considered for each diagram are shown in 
Tab.~\ref{tb1}.
%%%%%%%%%%%%%%%%%%%%%%%%%%%%%%%%%%%%%%%%%%%%%%%%%%%%%%%
\begin{table}[!htb]
\begin{ruledtabular}
  \begin{tabular}{lccc} 
 & $J^\pi$ & $I$ &$Mass$ ($MeV$) \\ \hline
$N$ & $1/2^+$&1/2&938\\ 
$N(1650)$& $1/2^-$&1/2&1950\\ 
$N(1710)$ &$1/2^+$&1/2 &1710\\ 
$N^*(1875)$& $3/2^-$&1/2 &1875\\ 
 $N^*(1900)$&$3/2^+$&1/2&1900\\ 
$\Xi$ & $1/2^+$&1/2&1320\\ 
 $\Xi^*(1820)$& $3/2^-$&1/2&1820 \\
 \end{tabular}
 \caption{Particles considered in the $K\Lambda$ interaction}\label{tb1}
\end{ruledtabular}
\end{table}
%%%%%%%%%%%%%%%%%%%%%%%%%%%%%%%%%%%%%%%%%%%%%%%%%%%%%%%%%%%%%%%%%%

 For the calculation of the contribution of
particles with spin-1/2 ($N$ and $\Xi$) in the intermediate state
(FIG.~\ref{fig1}a and c ),
 the Lagrangian of interaction is (considering the necessary 
adaptations from eq.~(\ref{eq1})) 
\begin{equation}
\label{eq}
\mathcal{L}_{ \Lambda K B}= \frac{g_{\Lambda KB}}{2m_{\Lambda}}\big(\overline{B}\gamma_\mu\gamma_5\Lambda\big)\partial^\mu\phi + {\rm H.c.} \ ,
\end{equation}
where $\phi$ represents the kaon field, $B$ the intermediate baryon field, with mass $m_B$, 
and $\Lambda$, the hyperon field, with mass $m_\Lambda$.

Calculating the Feynman diagrams and comparing with eq.~(\ref{eq20}) we find the 
amplitudes for the $N$
 (spin-1/2) particles contribution
\begin{equation}
\label{eqc8}
A_N=\frac{g^2_{\Lambda KN}}{4m_\Lambda^2}(m_N+m_\Lambda)\bigg(\frac{s-m_\Lambda^2}{s-m_N^2}\bigg)\ ,
\end{equation}
\begin{equation}
\label{eqc9}
\hspace{-0,5cm}B_N=-\frac{g^2_{\Lambda KN}}{4m_\Lambda^2}\bigg[\frac{2m_\Lambda(m_\Lambda+m_N)+s-m_\Lambda^2}{s-m_N^2}\bigg]\ ,
\end{equation}
and for the $\Xi$ (spin-1/2) hyperon in the crossed diagram (FIG.~\ref{fig1}c) 
the contribution is
\begin{equation}
\label{eqc8}
A_\Xi=\frac{g^2_{\Lambda K\Xi}}{4m_\Lambda^2}(m_\Xi+m_\Lambda)\bigg(\frac{u-m_\Lambda^2}{u-m_\Xi^2}\bigg)\ ,
\end{equation}
\begin{equation}
\label{eqc9}
\hspace{-0,5cm}B_\Xi=\frac{g^2_{\Lambda K\Xi}}{4m_\Lambda^2}\bigg[\frac{2m_\Lambda(m_\Lambda+m_\Xi)+u-m_\Lambda^2}{u-m_\Xi^2}\bigg]\ ,
\end{equation}
where $u$ is a Mandelstam variable, defined in the appendix and 
$g_{\Lambda KN(\Xi)}$ are the coupling constants.

In a similar way, we adapted the interaction Lagrangian (\ref{eq2}) for 
the exchange of spin-3/2 resonances, shown in FIG.~\ref{fig1}b and d
\begin{equation}
\label{eq}
\mathcal{L}_{ \Lambda KB^*}=  g_{ \Lambda KB^*}\Big\{\overline{B}^{*\mu}\big[g_{\mu\nu}-(Z+\frac{1}{2})\gamma_\mu\gamma_\nu\big]\Lambda\Big\}\partial^\nu\phi  +{\rm H.c.}\ .
\end{equation}

\noindent 
Calculating the amplitude for the exchange of a spin-3/2
$N^*$ (FIG.~\ref{fig1}b) we  have
\begin{equation}
A_{N^*}=\frac{g_{\Lambda KN^*}^2}{6}\bigg[\frac{2\hat{A}+3(m_\Lambda+m_{N^*})t}{m_{N^*}^2-s}+a_0\bigg]\ ,
\end{equation}
\begin{equation}
B_{N^*}=\frac{g_{\Lambda KN^*}^2}{6}\bigg[\frac{2\hat{B}+3t}{m_{N^*}^2-s}-b_0\bigg]\ ,
\end{equation}
where 
\begin{eqnarray}
\label{eq29'}
\hat{A}&=&3(m_\Lambda+m_{N^*})(q _{N^*})^2\nonumber\\
&&+(m_{N^*}-m_\Lambda)(E_{N^*}+m_\Lambda)^2\ ,
\end{eqnarray}
\begin{equation}
\label{eq30'}
\hat{B}=3(q_{N^*})^2-(E_{N^*}+m_\Lambda)^2\ ,
\end{equation}
\begin{eqnarray}
\label{eq31'}
a_0&=&-\frac{(m_\Lambda+m_{N^*})}{m_{N^*}^2}\Big(2m_{N^*}^2+m_\Lambda m_{N^*}\nonumber\\
&&-m_\Lambda ^2+2m_K^2\Big)+\frac{4}{m_{N^*}^2}\Big[(m_{N^*}+m_\Lambda )Z\nonumber\\
&&+(2m_{N^*}+m_\Lambda )Z^2\Big]\Big[s-m_\Lambda^2\Big]\ ,
\end{eqnarray}
\begin{eqnarray}
\label{eq32'}
b_0&=&\frac{8}{m_{N^*}^2}\Big[(m_\Lambda ^2+m_\Lambda m_{N^*}-m_K^2)Z\nonumber\\
&&+(2m_\Lambda m_{N^*}+m_\Lambda ^2)Z^2\Big]+\frac{(m_\Lambda +m_{N^*})^2}{m_{N^*}^2}\nonumber\\
&&+\frac{4Z^2}{m_{N^*}^2}\Big[s-m_\Lambda^2\Big]\ .
\label{eq:}
\end{eqnarray}
For the spin-3/2  $\Xi^*$ resonance (FIG.~\ref{fig1}d), the amplitudes are
\begin{eqnarray}
A_{\Xi^*}&=&\frac{g_{\Lambda K\Xi^*}^2}{6}\bigg[\frac{2\hat{A'}+3(m_\Lambda+m_{\Xi^*})t}{m_{\Xi^*}^2-u}\nonumber\\
&&+c_0+c_z(u-m_\Lambda^2)\bigg]\ ,
\end{eqnarray}
\begin{eqnarray}
B_{\Xi^*}&=&\frac{g_{\Lambda K\Xi^*}^2}{6}\bigg[-\frac{2\hat{B}'+3t}{m_{\Xi^*}^2-u}\nonumber\\
&&+d_0+d_z(u-m_\Lambda^2)\bigg]\ ,
\end{eqnarray}
where
\begin{eqnarray}
c_0&=&-\frac{(m_\Lambda+m_{\Xi^*})}{m_{\Xi^*}^2}(2m_{\Xi^*}^2+m_\Lambda m_{\Xi^*}\nonumber\\
&&-m_\Lambda ^2+2m_K^2)\ ,\\
c_z&=&\frac{4}{m_{\Xi^*}^2}\Big[(m_{\Xi^*}+m_\Lambda )Z\nonumber\\
&&+(2m_{\Xi^*}+m_\Lambda )Z^2\Big]\ ,
\end{eqnarray}
\begin{eqnarray}
d_0&=&\frac{8}{m_{\Xi^*}^2}\Big[(m_\Lambda ^2+m_\Lambda m_{\Xi^*}-m_K^2)Z+(2m_\Lambda m_{\Xi^*}+m_\Lambda ^2)Z^2\Big]\nonumber\\
&&+\frac{(m_\Lambda +m_{\Xi^*})^2}{m_{\Xi^*}^2}\ ,\\
d_z&=&\frac{4Z^2}{m_{\Xi^*}^2}\ ,
\label{eq:}
\end{eqnarray}
 where $t$, $q_{N^*}$, $E_{N^*}$ are defined in the appendix and $m_K$, $m_N*$ are the kaon and the $N^*$ masses respectively. For $\hat{A}'$ and $\hat{B}'$ we  change 
the $N^*$ parameters, inserting the $\Xi^*$ ones  in eqs.~(\ref{eq29'}) and 
(\ref{eq30'}). 
$g_{\Lambda KN^*(\Xi^*)}$ are the coupling constants.

For the last diagram, FIG.~$\ref{fig1}$e, the scalar $\sigma$ meson exchange,
 a parametrization of the amplitude has been considered \cite{BH}-\cite{ccb}  
\begin{equation}
A_\sigma=a+bt  \  ,
\label{eq32}
\end{equation}
\begin{equation}
B_\sigma=0   \  ,
\label{eq33}
\end{equation}
with $a=1,05 m_\pi^{-1}$, $b=-0,8m_\pi^{-3}$ and the pion mass $m_\pi$.
Some discussions about this term may be found in \cite{cm}, \cite{leut1}-\cite{r1}.

The parameters considered in the $K\Lambda $ interaction are shown in Tab. \ref{tb2}, the masses are taken from \cite{pdg}.
 
%%%%%%%%%%%%%%%%%%%%%%%%%%%%%%%%%%%%%%%%%%%%%%%%
\begin{table}[!htb]
\begin{ruledtabular}
  \begin{tabular}{ll}  
$m_\pi$& 140 $MeV$ \\
$m_K$& 496 $MeV$ \\ 
$m_\Lambda$& 1116 $MeV$ \\ 
$Z$& $-0,5$ \\ 
$g_{\Lambda KN}$ & 11,50\\
$g_{\Lambda KN(1650)}$ & 10,7 $GeV^{-1}$\\
$g_{\Lambda KN(1710)}$ & 5,2 $GeV^{-1}$\\
$g_{\Lambda KN^*(1875)}$& 0,53 $GeV^{-1}$ \\
$g_{\Lambda KN^*(1900)}$ & 2,6 $GeV^{-1}$\\
$g_{\Lambda K\Xi}$ & 0,2 \\
$g_{\Lambda K\Xi^*(1820)}$& 1,8 $GeV^{-1}$\\  
 \end{tabular}
 \caption{Parameters for the $K\Lambda$ interaction}\label{tb2}
\end{ruledtabular}
\end{table}
%%%%%%%%%%%%%%%%%%%%%%%%%%%%%%%%%%%%%%%%%%%%%%%%

The coupling constants $g_{\Lambda KN}$ and $g_{\Lambda K\Xi}$  are determined using $SU(3)$ \cite{stri, swart} and the ones of the $\Lambda$ with the resonances, 
by using the Breit-Wigner expresion eq. ($\ref{eq19}$),
as described above, in the same way it has been done in \cite{BH}. 

\begin{figure}[!htb]
 \includegraphics[width=0.5\textwidth]{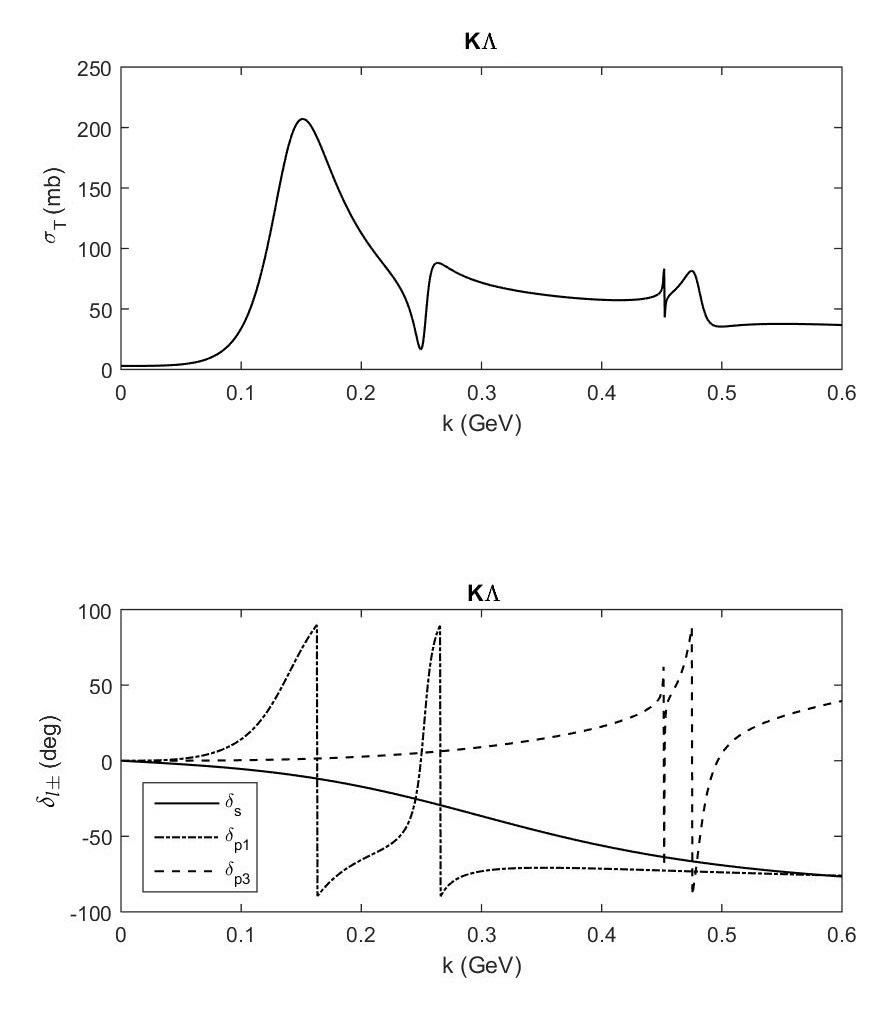}
 \caption{Total Cross Section and Phase Shifts of the $K\Lambda$ scattering}\label{fig2}
\end{figure}

\begin{figure} 
\includegraphics[width=0.5\textwidth]{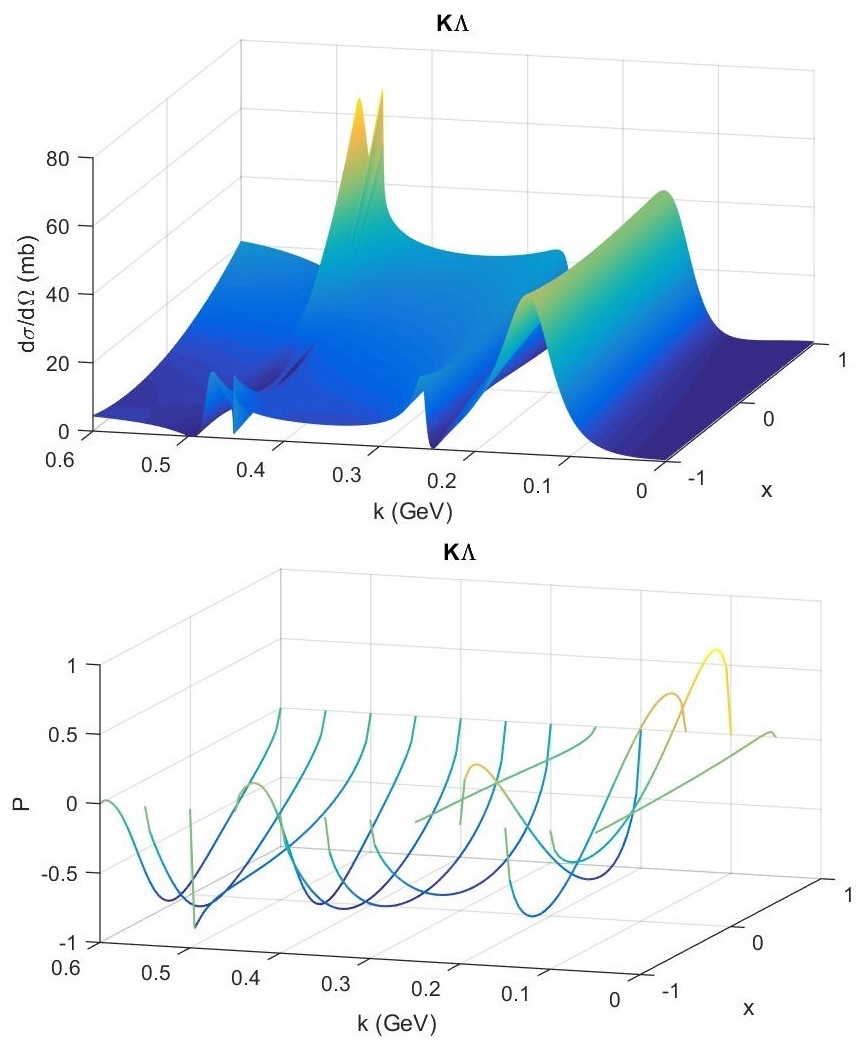}
\caption{Angular Distribution and Polarization of the $K\Lambda$ scattering}\label{fig3}
\end{figure}

In FIG.~\ref{fig2}, we show our results for the total elastic cross section 
and the phase shifts as functions of the kaon momentum $k$,
defined  in the center-of-mass frame. 
 Figure FIG.~\ref{fig3} shows the angular distributions and the polarizations 
as functions of $x=\cos\theta$ and $k$.


Observing the figure
we can note that the resonances, and in special the $N(1650)$ contribution, dominate the total
 cross section when $k\sim$ 150 MeV, as it was expected. At higher energies, 
 the other resonances also have an important effect. The polarization oscilates for
 $k<150$ MeV, but as the momentum increases, it becomes negative.

%%%%%%%%%%%%%%%%%%%%%%%%%%%%%%%%%%%%%%%%%%%%%%%%%%%%%%%%%%%%%%%%%%%%%%%%%%%%%%%%%%%%%%%%%%%%%%%%%%%%%%%%%%%%%%%%%%%%%%%%%%%%%%%%%%%%
\section{Antikaon-Lambda Interaction}

\begin{figure}[!htb]
 \includegraphics[width=0.5\textwidth]{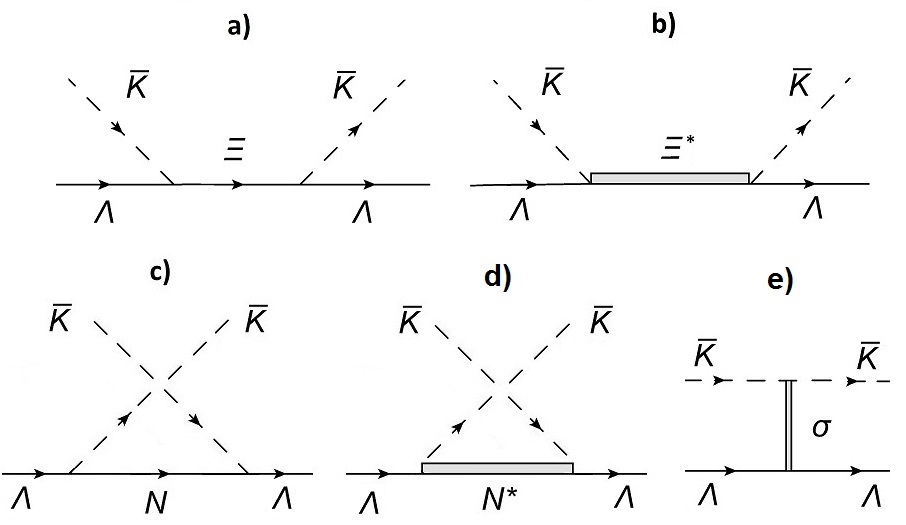}
 \caption{Diagrams for the $\overline K\Lambda$ interaction}\label{fig4} 
\end{figure}

The $\overline K\Lambda$ interations may be studied
 exactly in same way as it has been done in the last section for the
 the $K\Lambda$ interactions. 
Now we have the contributions presented in FIG.~\ref{fig4}, where the Lagrangians
 take into account the $N$, $\Xi$, $\Lambda$ and $\phi'$ 
(representing the antikaon) fields

\begin{equation}
\label{eq}
\mathcal{L}_{ \Lambda \overline{K} B}= \frac{g_{\Lambda \overline{K} B}}{2m_{\Lambda}}\big(\overline{B}\gamma_\mu\gamma_5\Lambda\big)\partial^\mu\phi'\ ,
\end{equation}
\begin{equation}
\label{eq}
\mathcal{L}_{ \Lambda \overline{K}B^*}=  g_{ \Lambda \overline{K}B^*}\Big\{\overline{B}^{*\mu}\big[g_{\mu\nu}-(Z+\frac{1}{2})\gamma_\mu\gamma_\nu\big]\Lambda\Big\}\partial^\nu\phi'\ .
\end{equation}

\begin{figure}[!htb]
\includegraphics[width=0.5\textwidth]{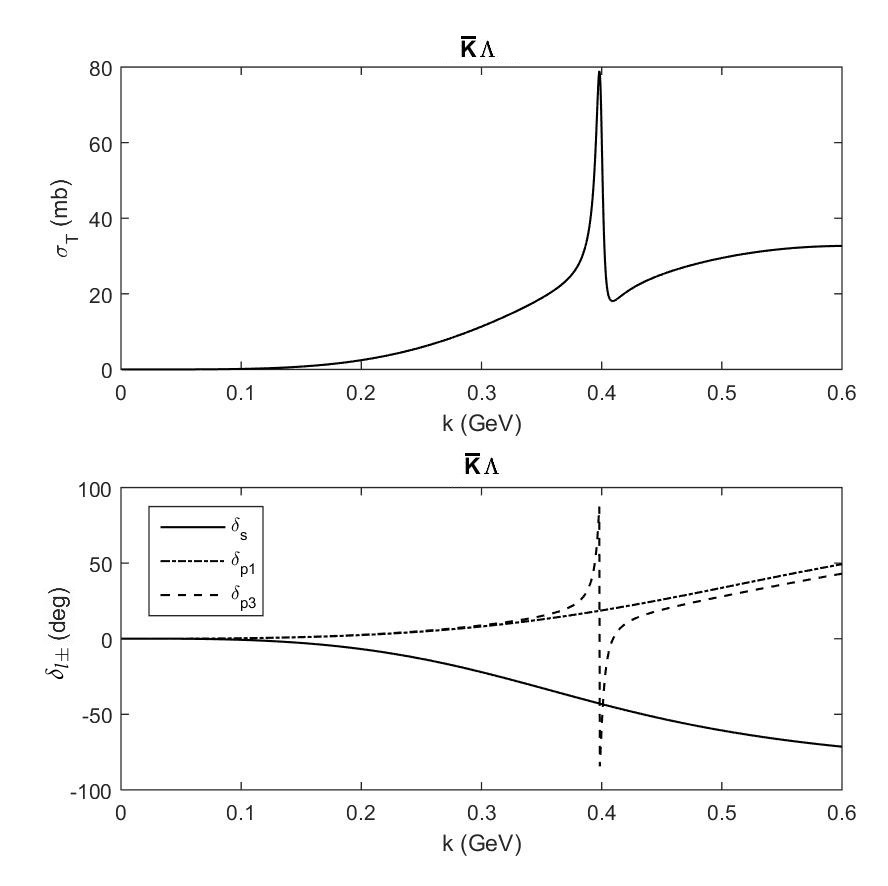}
 \caption{Total Cross Section and Phase Shifts of 
 the $\overline K\Lambda$ scattering}\label{fig5}
\includegraphics[width=0.5\textwidth]{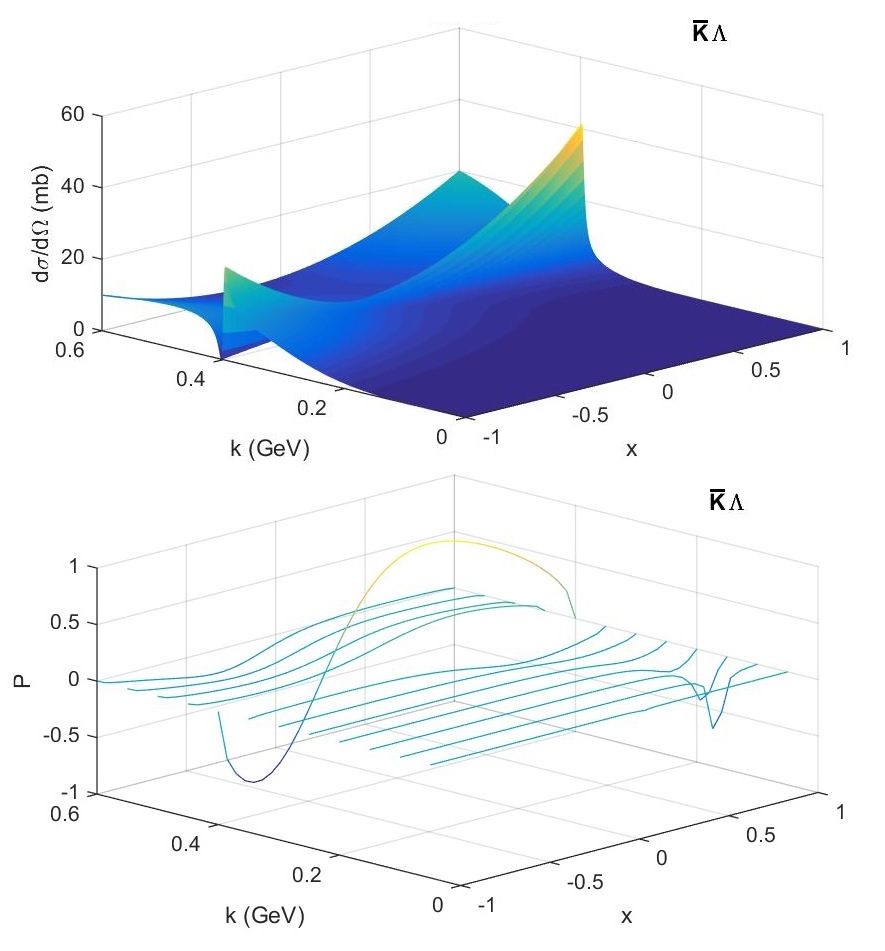}
\caption{Angular Distribution and Polarization in the $\overline{K}\Lambda$ scattering}\label{fig6}
\end{figure}
\noindent
The parameters considered are given before,
 $m_{\overline{K}} = m_K$, and for the crossed diagrams in FIG.~\ref{fig4}c and d
 we have considered only $N$(938) and $N^*$(1900), that are the most important processes.
The amplitudes ($\ref{eq32}$) and ($\ref{eq33}$)
have been calculated and the results are shown 
 in Figures $\ref{fig5}$ and $\ref{fig6}$.

%%%%%%%%%%%%%%%%%%%%%%%%%%%%%%%%%%%%%%%%%%%%%%%%%%%%%%%%%%%%%%%%%%%%%%%%%%%%%%%%%%%%%%%%%%%%%%%%%%%%%%%%%%%%%%%%%%%%%%%%%%%%%%%%%%%%
 
\section{Kaon-Sigma Interaction}
In this case the interacting particles  have 
isospin 1/2 and 1 ($K$ and  $\Sigma$ respectively). So, we have two possible total 
isospin states, 1/2 and 3/2, which allow also the exchange of $\Delta$ particles.
 
The scattering amplitude has the general form
\begin{figure}[!htb]
 \includegraphics[width=0.50\textwidth]{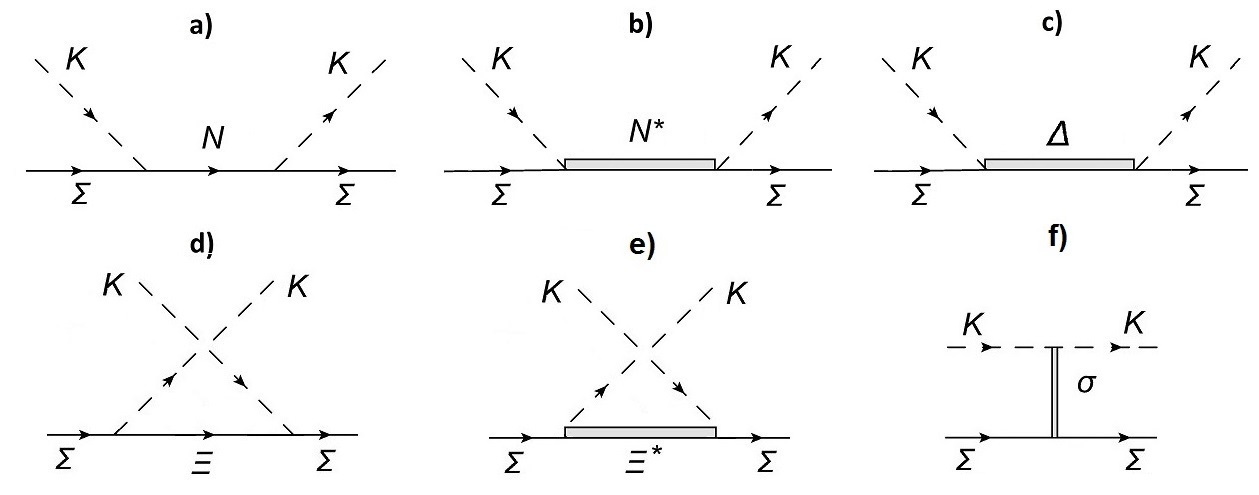}
 \caption{Diagrams for the $K\Sigma$ interaction}\label{fig7} 
\end{figure}
\[
T_{K\Sigma}^{\beta\alpha}=\bar{u}(\vec{p'})\bigg\{\Big[A^++\Big(\frac{\slashed k+\slashed k'}{2}\Big)B^+\Big]\delta^{\beta\alpha}
\]
\begin{equation}
+\Big[A^-+\Big(\frac{\slashed k+\slashed k'}{2}\Big)B^-\Big]i\epsilon^{\beta\alpha c}\tau_c\bigg\}u(\vec{p})\ ,
\label{eq}
\end{equation}  
and the considered projection operators are
\begin{equation}
P^{\beta\alpha}_{\frac{1}{2}}=\frac{1}{3}\delta^{\beta\alpha}+\frac{i}{3}\epsilon^{\beta\alpha c}\tau_c \ ,
\label{eq37}
\end{equation}
\begin{equation}
P^{\beta\alpha}_{\frac{3}{2}}=\frac{2}{3}\delta^{\beta\alpha}-\frac{i}{3}\epsilon^{\beta\alpha c}\tau_c\ ,
\label{eq38}
\end{equation}
where the indices $\alpha$ and $\beta$ are relative to
the initial and final isospin states of the $\Sigma$. 
%%%%%%%%%%%%%%%%%%%%%%%%%%%%%%%%%%%%%%%%%%%%%%%%%%%%%%%
\begin{table}[!htb]
\begin{ruledtabular}
  \begin{tabular}{lccc}  
 & $J^\pi$ & $I$ &$Mass$ ($MeV$) \\ \hline
$N$ & $1/2^+$&1/2&938\\ 
$N(1710)$ &$1/2^+$&1/2 &1710\\ 
$N^*(1875)$& $3/2^-$&1/2 &1875\\ 
 $N^*(1900)$&$3/2^+$&1/2&1900\\ 
$\Delta$& $3/2^+$&3/2&1920\\ 
$\Xi$ & $1/2^+$&1/2&1320\\ 
 $\Xi^*(1820)$& $3/2^-$&1/2&1820 \\
 \end{tabular}
 \caption{Resonances of the $K\Sigma$ interaction }\label{tb3} 
\end{ruledtabular}
\end{table}
%%%%%%%%%%%%%%%%%%%%%%%%%%%%%%%%%%%%%%%%%%%%%%%%%%%%%%%%%%%%%%%%%%

The contributing diagrams are shown
in FIG.~\ref{fig7} and the considered particles
 in Tab.~\ref{tb3}. The Lagrangians  ($\ref{eq1}$), ($\ref{eq2}$) now become,
\begin{equation}
\label{eq39}
\mathcal{L}_{\Sigma K B}= \frac{g_{\Sigma KB}}{2m_{\Sigma}}\big(\overline{B}\gamma_\mu\gamma_5\vec{\tau}.\vec{\Sigma}\big)\partial^\mu\phi  \  ,
\end{equation}
\begin{equation}
\label{eq40}
\mathcal{L}_{\Sigma KB^*}=  g_{ \Sigma KB^*}\Big\{\overline{B}^{*\mu}\big[g_{\mu\nu}-(Z+\frac{1}{2})\gamma_\mu\gamma_\nu\big]\vec{Q}.\vec{\Sigma}\Big\}\partial^\nu\phi  \  ,
\end{equation}
\noindent
where $\vec{Q}$ is the $\vec{M}$ matrix for $\Delta$ ($I=3/2$) or $\vec{\tau}$ matrix for the $N^*$ and $\Xi^*$ ($I=1/2$).

The resulting amplitudes for nucleons in the intermediate state
 (FIG.~\ref{fig7}a) are
\begin{equation}
\label{eq50'}
A_N^+=\frac{g^2_{\Sigma KN}}{4m_\Sigma^2}(m_N+m_\Sigma)\bigg(\frac{s-m_\Sigma^2}{s-m_N^2}\bigg)\ ,
\end{equation}
\begin{equation}
\label{eqc9}
\hspace{-0,8cm}B_N^+=-\frac{g^2_{\Sigma KN}}{4m_\Sigma^2}\bigg[\frac{2m_\Sigma(m_\Sigma+m_N)+s-m_\Sigma^2}{s-m_N^2}\bigg]\ ,
\end{equation}
\begin{equation}
\label{eqc10}
A_N^-=\frac{g^2_{\Sigma KN}}{4m_\Sigma^2}(m_N+m_\Sigma)\bigg(\frac{s-m_\Sigma^2}{s-m_N^2}\bigg)\ ,
\end{equation}
\begin{equation}
\label{eqc11}
\hspace{-0,8cm}B_N^-=-\frac{g^2_{\Sigma KN}}{4m_\Sigma^2}\bigg[\frac{2m_\Sigma(m_\Sigma+m_N)+s-m_\Sigma^2}{s-m_N^2}\bigg]\ ,
\end{equation}
and for the $\Xi$ exchange in the diagram \ref{fig7}d  
\begin{equation}
\label{eqc8}
A_\Xi^+=\frac{g^2_{\Sigma K\Xi}}{4m_\Sigma^2}(m_\Xi+m_\Sigma)\bigg(\frac{u-m_\Sigma^2}{u-m_\Xi^2}\bigg)\ ,
\end{equation}
\begin{equation}
\label{eqc9}
\hspace{-0,8cm}B_\Xi^+=\frac{g^2_{\Sigma K\Xi}}{4m_\Sigma^2}\bigg[\frac{2m_\Sigma(m_\Sigma+m_\Xi)+u-m_\Sigma^2}{u-m_\Xi^2}\bigg]\ ,
\end{equation}
\begin{equation}
\label{eqc10}
A_\Xi^-=-\frac{g^2_{\Sigma K\Xi}}{4m_\Sigma^2}(m_\Xi+m_\Sigma)\bigg(\frac{u-m_\Sigma^2}{u-m_\Xi^2}\bigg)\ ,
\end{equation}
\begin{equation}
\label{eqc11}
\hspace{-0,8cm}B_\Xi^-=-\frac{g^2_{\Sigma K\Xi}}{4m_\Sigma^2}\bigg[\frac{2m_\Sigma(m_\Sigma+m_\Xi)+u-m_\Sigma^2}{u-m_\Xi^2}\bigg]\ .
\end{equation}
\noindent
Figure $\ref{fig7}$b gives
\begin{equation}
\label{eqcc77}
\hspace{-0,8cm}A_{N^*}^+=\frac{g_{\Sigma KN^*}^2}{6}\bigg[\frac{2\hat{A}+3(m_\Sigma+m_{N^*})t}{m_{N^*}^2-s}+a_0\bigg]\ ,
\end{equation}
\begin{equation}
\label{eqcc78}
B_{N^*}^+=\frac{g_{\Sigma KN^*}^2}{6}\bigg[\frac{2\hat{B}+3t}{m_{N^*}^2-s}-b_0\bigg]\ ,
\end{equation}
\begin{equation}
\label{eqcc79}
\hspace{-0,8cm}A_{N^*}^-=\frac{g_{\Sigma KN^*}^2}{6}\bigg[\frac{2\hat{A}+3(m_\Sigma+m_{N^*})t}{m_{N^*}^2-s}+a_0\bigg]\ ,
\end{equation}
\begin{equation}
\label{eqcc80}
B_{N^*}^-=\frac{g_{\Sigma KN^*}^2}{6}\bigg[\frac{2\hat{B}+3t}{m_{N^*}^2-s}-b_0\bigg]\ ,
\end{equation}
and for the crossed diagram shown in FIG.~\ref{fig7}e, where
a $\Xi^*$ exchange is taken into account,
\begin{eqnarray}
\label{eqcc77}
A_{\Xi^*}^+&=&\frac{g_{\Sigma K\Xi^*}^2}{6}\bigg[\frac{2\hat{A'}+3(m_\Sigma+m_{\Xi^*})t}{m_{\Xi^*}^2-u}\nonumber\\
&&+c_0+c_z(u-m_\Sigma^2)\bigg]\ ,
\end{eqnarray}
\begin{equation}
\label{eqcc78}
\hspace{-0,8cm}B_{\Xi^*}^+=\frac{g_{\Sigma K\Xi^*}^2}{6}\bigg[\frac{2\hat{B}'+3t}{m_{\Xi^*}^2-u}-d_0-d_z(u-m_\Sigma^2)\bigg]\ ,
\end{equation}
\begin{eqnarray}
\label{eqcc79}
A_{\Xi^*}^-&=&-\frac{g_{\Sigma K\Xi^*}^2}{6}\bigg[\frac{2\hat{A'}+3(m_\Sigma+m_{\Xi^*})t}{m_{\Xi^*}^2-u}\nonumber\\
&&+c_0+c_z(u-m_\Sigma^2)\bigg]\ ,
\end{eqnarray}
\begin{equation}
\label{eq65'}
\hspace{-0,7cm}B_{\Xi^*}^-=\frac{g_{\Sigma K\Xi^*}^2}{6}\bigg[\frac{2\hat{B}'+3t}{m_{\Xi^*}^2-u}-d_0-d_z(u-m_\Sigma^2)\bigg]\ ,
\end{equation}
where the expressions for $\hat{A}$, $\hat{B}$, $\hat{A}'$, $\hat{B}'$, $a_0$, $b_0$, $c_0$, $d_0$, $c_z$ and $d_z$ are the same ones presented in Sec. III,
but replacing the $\Lambda$ hyperon for the $\Sigma$ hyperon.

For the spin-isospin-3/2 $\Delta$ resonance in FIG.~\ref{fig4}c, we have the amplitudes 
\begin{equation}
\label{cc47}
\hspace{-0,7cm}A_\Delta^{+}=\frac{g_{\Sigma K\Delta}^2}{9}\Bigg\{\bigg[\frac{2\hat{A''}+3(m_\Sigma+m_\Delta)t}{m_\Delta^2-s}\bigg]
+a_0''\Bigg\}\ ,
\end{equation} 
\begin{equation}
\label{cc48}
B_\Delta^{+}=\frac{g_{\Sigma K\Delta}^2}{9}\Bigg\{\bigg[\frac{2\hat{B''}+3t}{m_\Delta^2-s}\bigg]-b_0''\Bigg\}\ ,
\end{equation} 
\begin{equation}
\hspace{-0,7cm}A_\Delta^{-}=\frac{g_{\Sigma K\Delta}^2}{18}\Bigg\{\bigg[\frac{2\hat{A''}+3(m_\Sigma+m_\Delta)t}{m_\Delta^2-s}\bigg]
+a_0''\Bigg\}\ ,
\end{equation} 
\begin{equation}
\label{cc49}
B_\Delta^{-}=\frac{g_{\Sigma K\Delta}^2}{18}\Bigg\{\bigg[\frac{2\hat{B''}+3t}{m_\Delta^2-s}\bigg]-b_0''\Bigg\}\ ,
\end{equation}
where the expressions for $\hat{A}''$, $\hat{B}''$, $a_0''$ and $b_0''$ are
given in (\ref{eq29'}), (\ref{eq30'}), (\ref{eq31'}) and (\ref{eq32'}) replacing $\Lambda$ for $\Sigma$ and $N^*$ for $\Delta$.

For the
 $\sigma$ exchange (FIG.~\ref{fig7}f) the parametrization from
eqs. ($\ref{eq32}$) and ($\ref{eq33}$) will be considered.

Thus, to calculate the observables for each reaction
 we use ($\ref{eq37}$) and ($\ref{eq38}$), resulting in the amplitudes 

\begin{equation}
\label{eq55}
A^{\frac{1}{2}}=A^++2A^-\ ,
\end{equation}
\begin{equation}
\label{eq56}
B^{\frac{1}{2}}=B^++2B^-\ ,
\end{equation}
\begin{equation}
\label{eq57}
A^{\frac{3}{2}}=A^+-A^-\ ,
\end{equation}
\begin{equation}
\label{eq58}
B^{\frac{3}{2}}=B^+-B^-\ ,
\end{equation}
 
\noindent
and the parameters are shown in Tabs. \ref{tb2} and \ref{tb4}. 

%%%%%%%%%%%%%%%%%%%%%%%%%%%%%%%%%%%%%%%%%%%%%%%%
\begin{table}[!htb]
\begin{ruledtabular}
  \begin{tabular}{ll}  
$m_\Sigma$& 1190$MeV$ \\ 
$g_{\Sigma KN}$ & 6,9\\ 
$g_{\Sigma KN(1710)}$ & 8,4$GeV^{-1}$\\
$g_{\Sigma KN^*(1875)}$& 0,7$GeV^{-1}$ \\
$g_{\Sigma KN^*(1900)}$ & 1,3$GeV^{-1}$\\
$g_{\Sigma K\Delta}$& 1,7$GeV^{-1}$\\
$g_{\Sigma K\Xi}$ & 13,4 \\ 
$g_{\Sigma K\Xi^*(1820)}$& 1,8$GeV^{-1}$\\ 
 \end{tabular}
 \caption{Parameters for the $K\Sigma$ interaction}\label{tb4}
\end{ruledtabular}
\end{table}
%%%%%%%%%%%%%%%%%%%%%%%%%%%%%%%%%%%%%%%%%%%%%%%%

To determine the coupling constants $g_{\Sigma KN}$, $g_{\Sigma K\Xi}$ and the
ones with
resonances we take into account the same arguments presented in Sec. III.

\begin{figure}[!htb]
\includegraphics[width=0.5\textwidth]{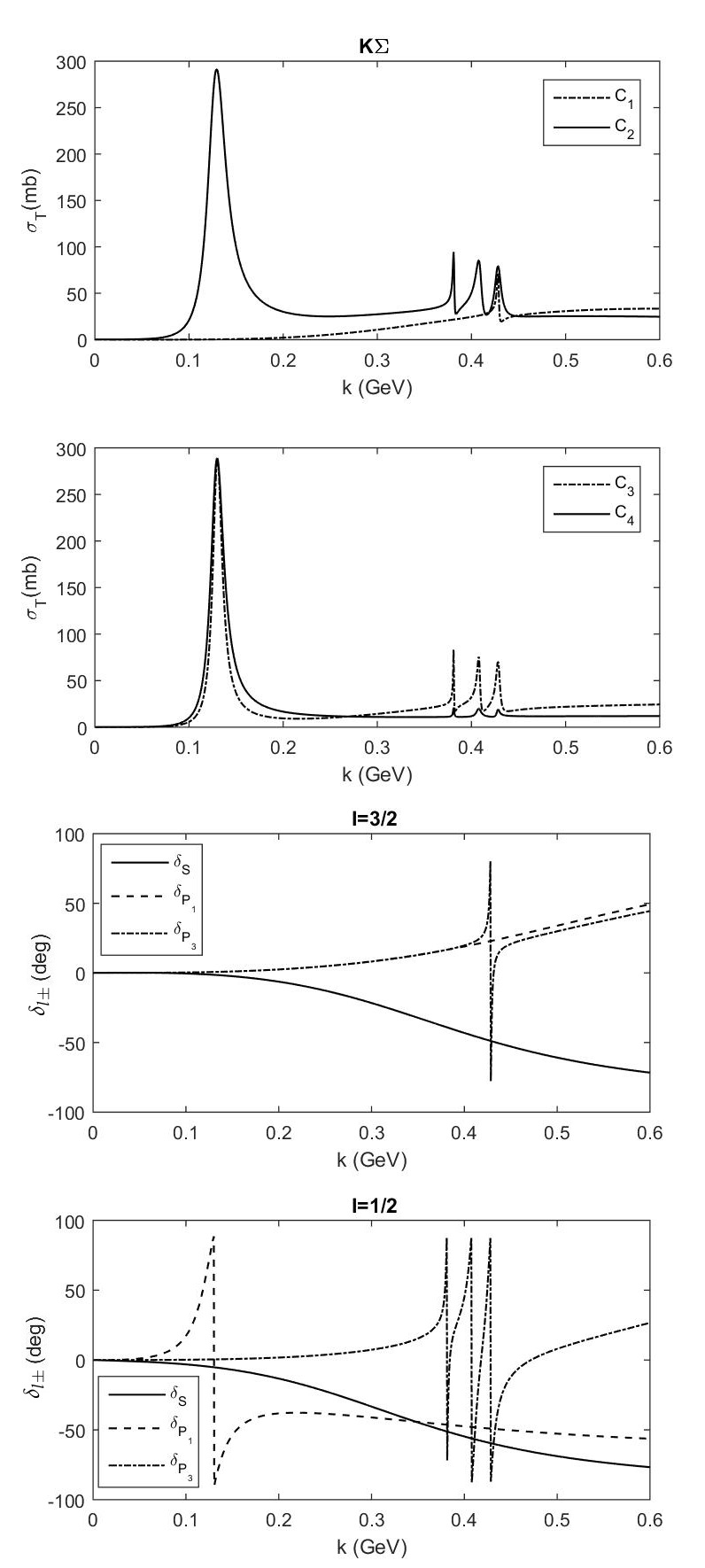}
 \caption{Total Cross Section and Phase Shifts of the $K\Sigma$ scattering}\label{fig8}
\end{figure}

\begin{figure}[!htb]
\includegraphics[width=0.5\textwidth]{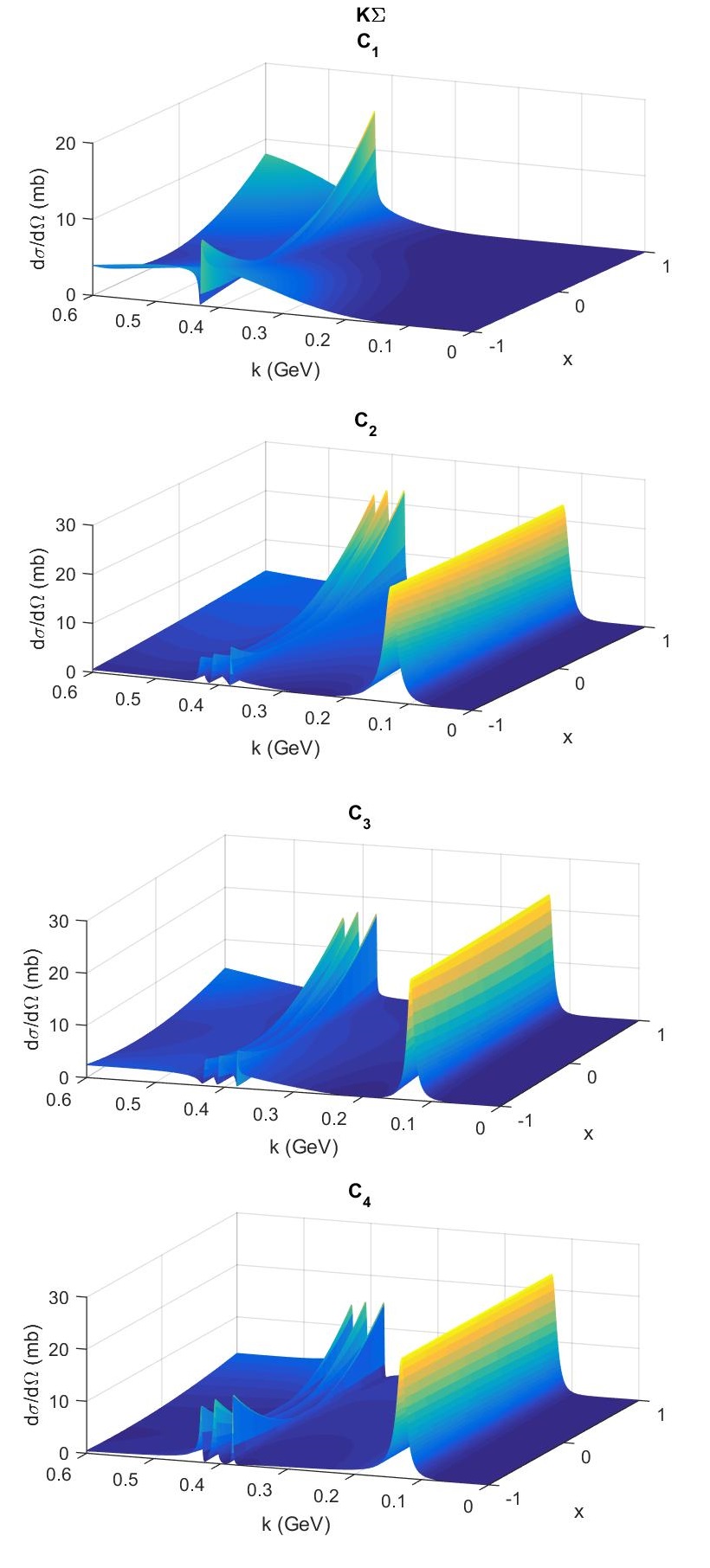}
\caption{Angular Distribution of the $K\Sigma$ scattering}\label{fig9}
\end{figure}

\begin{figure}[!htb]
\includegraphics[width=0.5\textwidth]{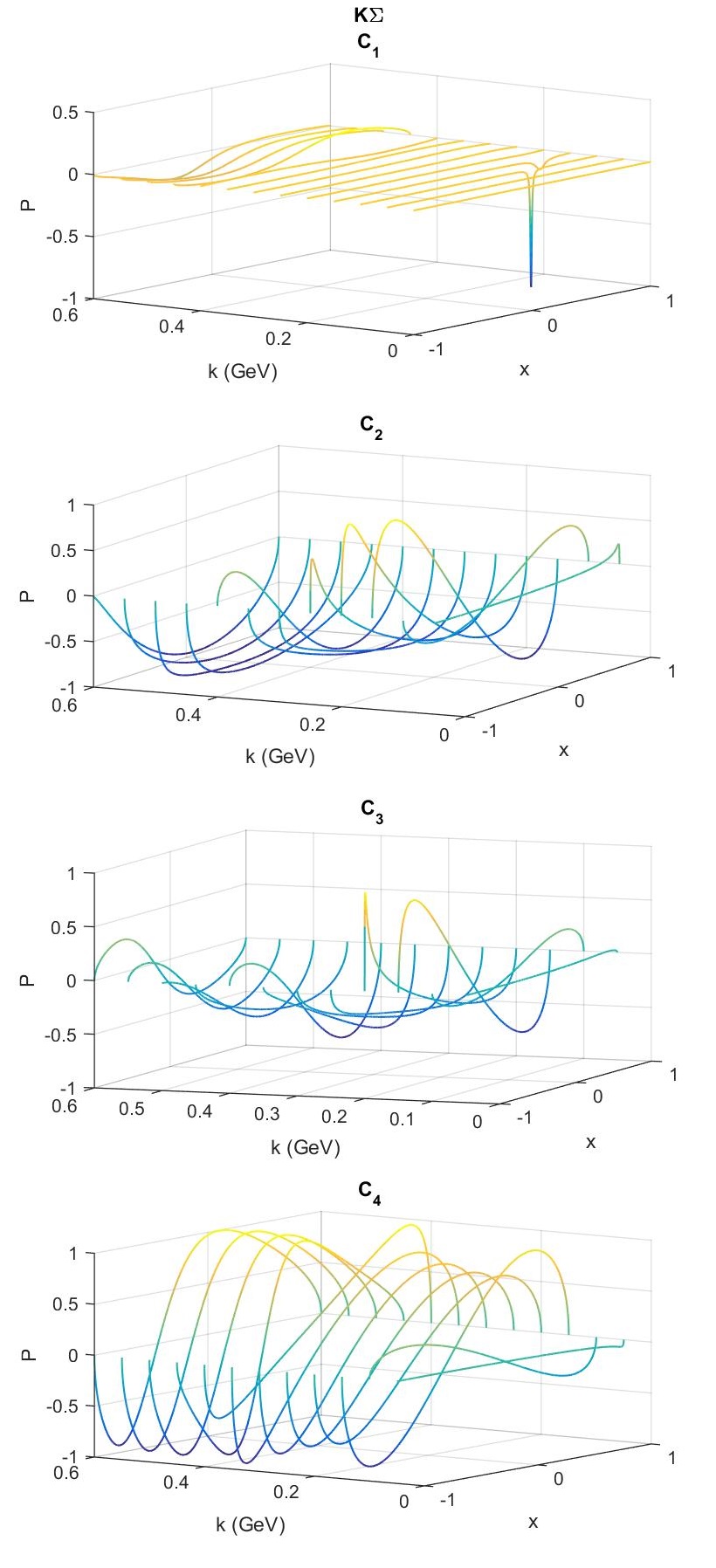}
\caption{Polarization in the $K\Sigma$ scattering}\label{fig9'}
\end{figure}

Using the isospin formalism for the elastic and the charge exchange scattering, 
we can determine the amplitudes for the reactions (that we name $C_i$, for simplicity)
\[
\left\langle\Sigma^+K^+|T|\Sigma^+K^+ \right\rangle=\left\langle\Sigma^-K^0|T|\Sigma^-K^0 \right\rangle
\]
\begin{equation}
=T_{\frac{3}{2}}\equiv C_1\ ,
\label{eq:}
\end{equation}

\[
\left\langle\Sigma^+K^0|T|\Sigma^+K^0 \right\rangle=\left\langle\Sigma^-K^+|T|\Sigma^-K^+ \right\rangle
\]
\begin{equation}
=\frac{1}{3}T_{\frac{3}{2}}+\frac{2}{3}T_{\frac{1}{2}}\equiv  C_2\ ,
\label{eq:}
\end{equation}
\[
\left\langle\Sigma^0K^0|T|\Sigma^0K^0 \right\rangle=\left\langle\Sigma^0K^+|T|\Sigma^0K^+ \right\rangle
\]
\begin{equation}
=\frac{2}{3}T_{\frac{3}{2}}+\frac{1}{3}T_{\frac{1}{2}}\equiv  C_3\ ,
\label{eq:}
\end{equation}
\[
\left\langle\Sigma^0K^0|T|\Sigma^-K^+ \right\rangle=\left\langle\Sigma^+K^0|T|\Sigma^0K^+ \right\rangle
\]
\[
=\left\langle\Sigma^-K^+|T|\Sigma^0K^0 \right\rangle=
\left\langle\Sigma^0K^+|T|\Sigma^+K^0 \right\rangle
\]
\begin{equation}
=\frac{\sqrt{2}}{3}\Big(T_{\frac{3}{2}}-T_{\frac{1}{2}}\Big)\equiv  C_4\ ,
\label{eq:}
\end{equation}
and
with these amplitudes we can calculate all observables of interest. 
The total elastic cross sections and the phase shifts as functions of the kaon momentum
are  shown in FIG.~\ref{fig8}. Figures $\ref{fig9}$ and   $\ref{fig9'}$ show the angular distributions and the 
polarizations.

%%%%%%%%%%%%%%%%%%%%%%%%%%%%%%%%%%%%%%%%%%%%%%%%%%%%%%%%%%%%%%%%%%%%%%%%%%%%%%%%%%%%%%%%%%%%%%%%%%%%%%%%%%%%%%%%%%%%%%%%%%%%%%%%%%%

\section{Antikaon-Sigma Interaction}

\begin{figure}[!htb]
 \includegraphics[width=0.5\textwidth]{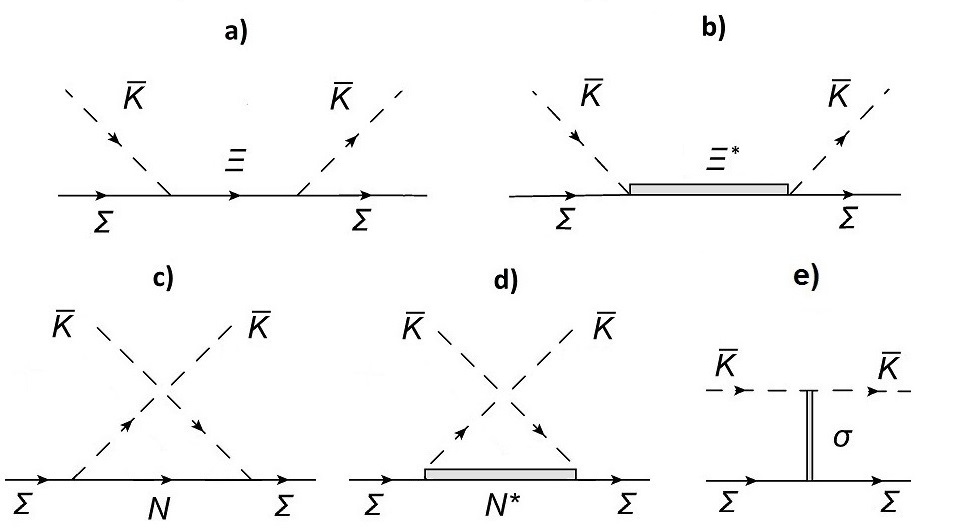}
 \caption{Diagrams for the $\overline K\Sigma$ interaction}\label{fig10}
\end{figure}

In this case, we will proceed in the same way as we have done
in the last section for the $K\Sigma$ interaction.  The diagrams to be considered are shown 
in FIG.~\ref{fig10}.

The  Lagrangians are very similar to the ones
used to study the
 $K\Sigma$ interaction,
$(\ref{eq39})$, $(\ref{eq40})$, 
replacing $N$ and $N^*$ for $\Xi$ and $\Xi^*$.
Then, if 
 these changes are implemented, we may
use the same amplitudes given by (\ref{eq50'})-(\ref{eq65'}),
 (\ref{eq55})-(\ref{eq58}). 

In this case we have the following reactions
\[
\left\langle \overline K^0\Sigma^+|T| \overline K^0\Sigma^+ \right\rangle=\left\langle K^-\Sigma^-|T|  K^-\Sigma^- \right\rangle
\]
\begin{equation}
=T_{\frac{3}{2}}\equiv D_1 \ ,
\end{equation}
\[
\left\langle\Sigma^+K^-|T|\Sigma^+K^- \right\rangle=\left\langle\Sigma^-\overline{K}^0|T|\Sigma^-\overline{K}^0 \right\rangle
\]
\begin{equation}
=\frac{1}{3}T_{\frac{3}{2}}+\frac{2}{3}T_{\frac{1}{2}}\equiv D_2\ ,
\label{eq:}
\end{equation}
\[
\left\langle\Sigma^0\overline{K}^0|T|\Sigma^0\overline{K}^0 \right\rangle=\left\langle\Sigma^0K^-|T|\Sigma^0K^- \right\rangle
\]
\begin{equation}
=\frac{2}{3}T_{\frac{3}{2}}+\frac{1}{3}T_{\frac{1}{2}}\equiv D_3\ ,
\label{eq:}
\end{equation}
\[
\left\langle\Sigma^0K^-|T|\Sigma^-\overline K^0 \right\rangle=\left\langle\Sigma^+K^-|T|\Sigma^0\overline K^0 \right\rangle
\]
\[
=\left\langle\Sigma^-\overline K^0|T|\Sigma^0K^- \right\rangle=
\left\langle\Sigma^0\overline K^0|T|\Sigma^+K^- \right\rangle
\]
\begin{equation}
=\frac{\sqrt{2}}{3}\Big(T_{\frac{3}{2}}-T_{\frac{1}{2}}\Big)\equiv  D_4\ .
\label{eq:}
\end{equation}

\begin{figure}[!htb]
\includegraphics[width=0.50\textwidth]{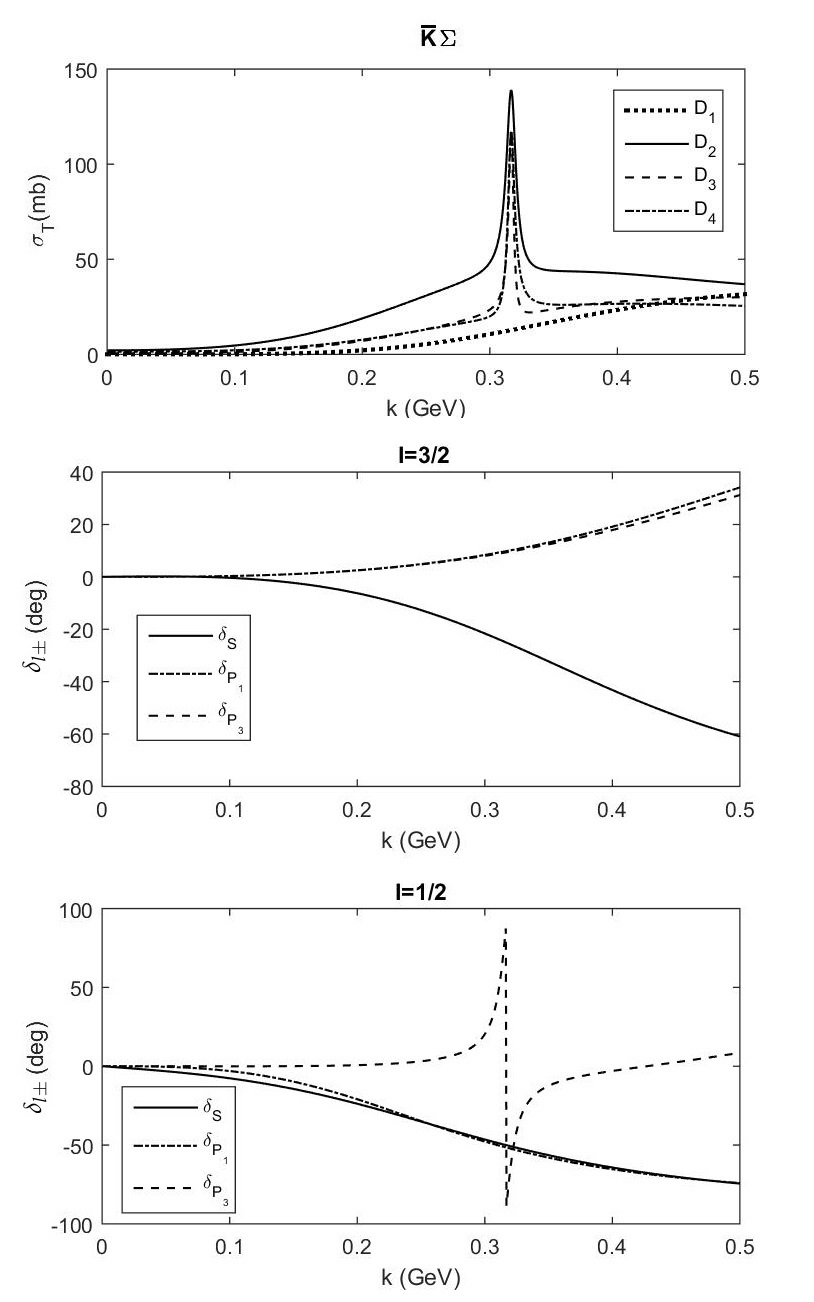}
 \caption{Total Cross Section and Phase Shifts of the 
$\overline K\Sigma$ scattering}\label{fig11}
\end{figure}

\begin{figure}[!htb]
\includegraphics[width=0.50\textwidth]{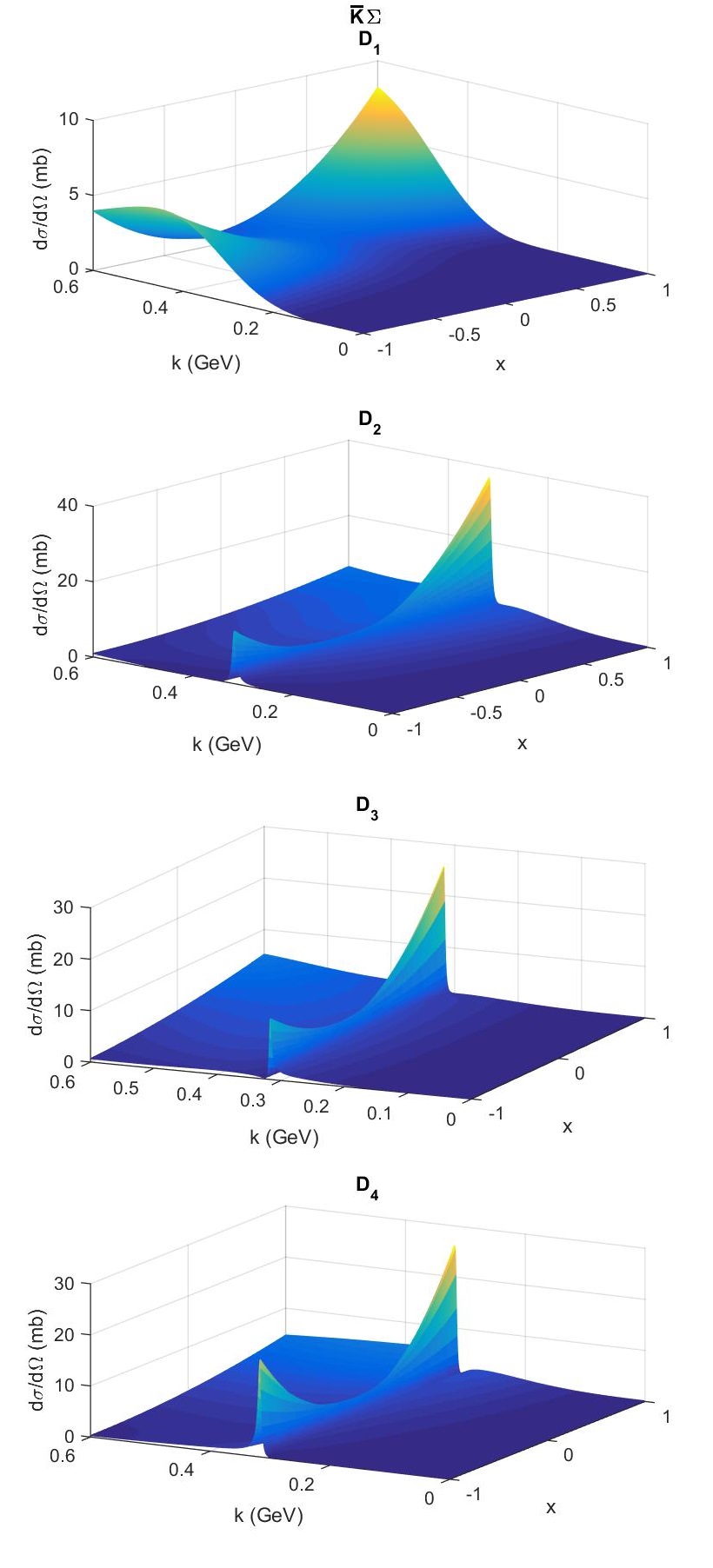}
\caption{Angular Distribution of the $\overline K\Sigma$ scattering}\label{fig12}
\end{figure}

\begin{figure}[!htb]
\includegraphics[width=0.50\textwidth]{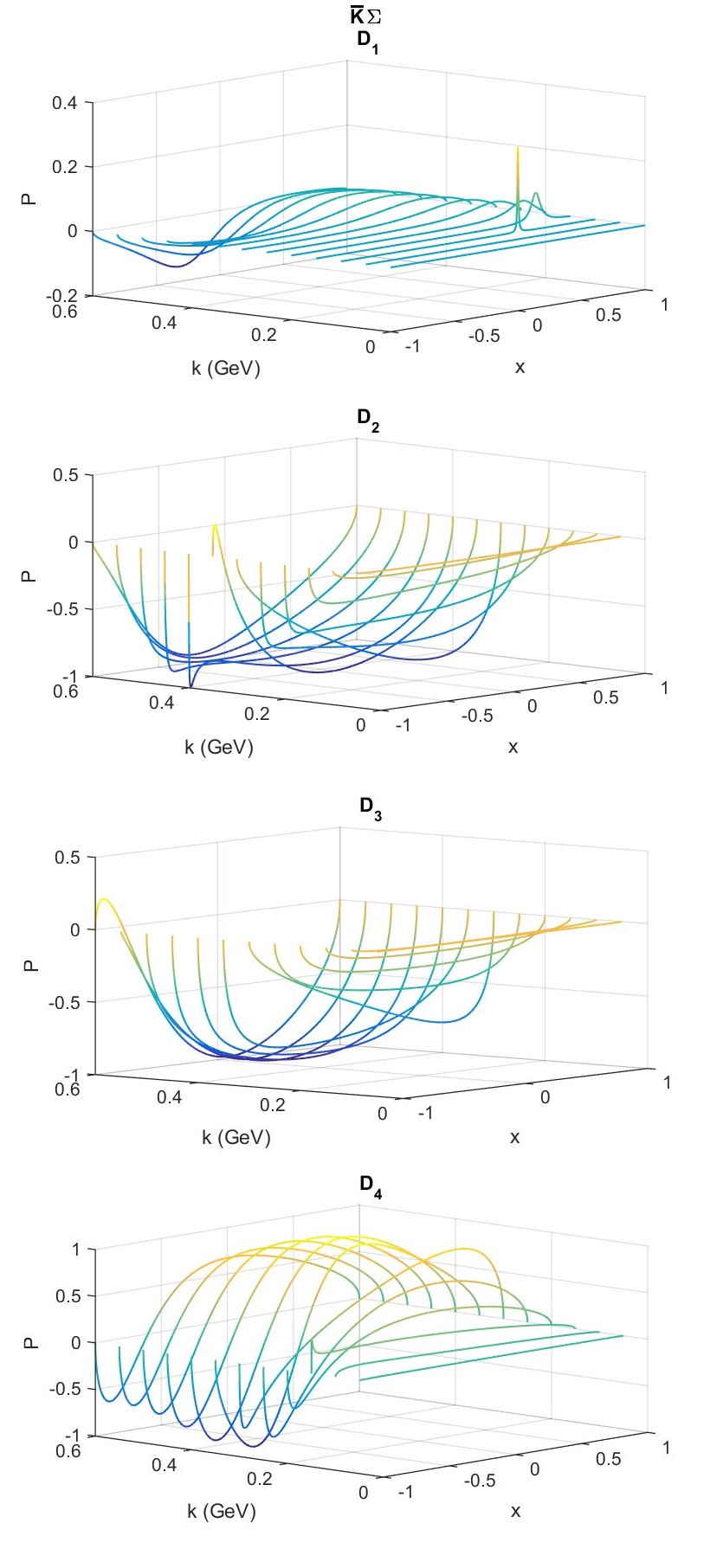}
\caption{Polarization in the $\overline K\Sigma$ scattering}\label{fig13}
\end{figure}

For diagram 11d the resonance to be considered is $N^*(1900)$.
Using the parameters given in Tab. \ref{tb4} 
we have obtained the results for the $\overline K\Sigma$ scattering shown in
 Figures $\ref{fig11}$, $\ref{fig12}$ and $\ref{fig13}$. 

%%%%%%%%%%%%%%%%%%%%%%%%%%%%%%%%%%%%%%%%%%%%%%%%%%%%%%%%%%%%%%%%%%%%%%%%%%%%%%%%%%%%%%%%%%%%%%%%%%%%%%%%%%%%%%%%%%%%%%%%%%%%%%%%%%%

\section{Discussion and Results}

In this work the low energy $K\Lambda$, ${\overline K}\Lambda$,
$K\Sigma$ and ${\overline K}\Sigma$ interactions have been studied
considering a model based on effective Lagrangians where  mesons, 
baryons and baryonic resonances have been taken into account. The coupling
constants have been determined and then the $S$ and $P$ phase shifts, cross
sections and polarizations have been calculated and shown in the figures 
of the previous sections. As it was expected,
for many channels the resonances dominated the cross sections, and for this reason,
we believe in the formulation of the proposed model at low energies ($k<$ 0.4 GeV).
In \cite{BH} a similar behavior has been observed, and the preditions of the model, 
when compared with the HyperCP data, showed to be very accurate.

For the ${\overline K}\Lambda$ scattering, 
at the $\Omega$ hyperon mass ($m_\Omega=$1672 MeV),
 $\delta_{P_1}=2.71^o$, $\delta_{P_3}=2.90^o$, $\delta_{D_3}=-0.0008^o$ and $\delta_{D_5}=-0.0001^o$. 
These strong phases may be used in a possible search of CP
violation in the $\Omega\rightarrow {\overline K}\Lambda$ decay,
in addition to the weak
CP violating phases, in the same way it has been done in \cite{ccb}
 (even considering that for other similar
decays, no CP violation has been observed \cite{hyper1}, \cite{hyper2}).

In the study of high energy hyperon polarization, 
produced in proton-nucleus and in heavy ion collisions,
if we consider the polarizations
obtained in the final-state interactions, the 
processes studied in this work may have some effect in the final polarization
of the $\Lambda$, ${\overline \Lambda}$, $\Sigma$ and ${\overline\Sigma}$ 
hyperons produced in
these reactions. In special, in some reactions of Fig. 14
considerable polarization may be observed, and some signs of this fact
 probably may be observed
in the $\Sigma$ and in the $\overline\Sigma$ polarizations. 
Probably this effect is smaller than the one obtained when
considering the $\pi Y$ interactions \cite{cy}-\cite{ccb2}, but as these 
interactions ($KY$)  provide polarizations of different signs, it is possible
to obtain some differences in the final results. And certainly, a more accurate
final result will be obtained.

These reactions are also important in the determination of nucleon-hyperon and 
hyperon-hyperon potentials, as they are subprocesses of these interactions, 
and as it has been discussed before, these interactions have a fundamental
importance in the structure of the hypernuclei and in the hyperon stars.

It must be pointed that the study of the $\Xi$ hyperons and other related reactions 
have been left for future
works.

So, as it has been shown, the study presented in this work is very important for
many physical systems of interest, and for this reason, must be continuated
and improved in
future works.

%%%%%%%%%%%%%%%%%%%%%%%%%%%%%%%%%%%%%%%%%%%%%%%%%%%%%%%%%%%%%%%%%
\section{Acknowledgments}

We would like to thank CAPES for the financial 
support.

%%%%%%%%%%%%%%%%%%%%%%%%%%%%%%%%%%%%%%%%%%%%%%%%%%%%%%%%%%%%%%%%%
\section{Appendix}
Considering a process where
 $p$ and $p'$ are
the initial and final hyperon four-momenta, $k$ and $k'$ are the initial and final meson 
four-momenta, the Mandelstam variables are given by

\begin{equation}
\label{eq:}
s=(p+k)^2=(p'+k')^2=m^2+m_K^2+2Ek_0-2\vec{k}.\vec{p}\ ,
\end{equation}
\begin{equation}
\label{eq:}
u=(p'-k)^2=(p-k')^2=m^2+m_K^2-2Ek_0-2\vec{k}'.\vec{p}\ ,
\end{equation}
\begin{equation}
\label{eq:}
t=(p-p')^2=(k-k')^2=2|\vec{k}|^2x-2|\vec{k}|^2  \   .
\end{equation}
In the center-of-mass frame, the energies will be
defined as
\begin{equation}
\label{eq:}
k_0=k'_0=\sqrt{|\vec{k}|^2+m_K^2}\ ,
\end{equation}
\begin{equation}
\label{eq:}
E=E'=\sqrt{|\vec{k}|^2+m^2}\ ,
\end{equation}
and the total momentum is null 
\begin{equation}
\label{eq:}
\vec{p}+\vec{k}=\vec{p}'+\vec{k}'=0\ .
\end{equation}
We also define the variable
\begin{equation}
\label{eq:}
x=\cos\theta\ ,
\end{equation}
where $\theta$ is the scattering angle.
Other variables of interest are
\begin{equation}
\label{eq:}
\nu_r=\frac{m_r^2-m^2-k.k'}{2m}\ ,
\end{equation}
\begin{equation}
\label{eq:}
\nu=\frac{s-u}{4m}=\frac{2Ek_0+|\vec{k}|^2+|\vec{k}|^2x}{2m}\ ,
\end{equation}

\begin{equation}
\label{eq:}
k.k'=m_K^2+|\vec{k}|^2-|\vec{k}|^2x=k_0^2-|\vec{k}|^2x\ ,
\end{equation}
where $m$, $m_r$ and $m_K$ are the hyperon mass, the resonance mass and the kaon mass, respectively.

 For
the energy and the 3-momentum of the intermediary particles we also
have the relations
\begin{equation}
(E_{B^*}\pm m_\Lambda)=\frac{(m_{B^*}\pm m_\Lambda)^2-m_K^2}{2m_{B^*}}\ ,
\end{equation}
\begin{equation}
(q_{B^*})^2=|\vec{q}_{B^*}|^2=E_{B^*}^2- m_\Lambda^2=(E_{B^*}+ m_\Lambda)(E_{B^*}- m_\Lambda)\ ,
\end{equation}
where $E_{B^*}$ and $\vec{q}_{B^*}$  are the energy and the momentum of  intermediary baryon $B^*$ in the center-of-mass frame, respectively.

%----------------------------------------------------------------------------------------
%	REFERENCE LIST
%----------------------------------------------------------------------------------------

%----------------------------------------------------------------------------------------


\begin{thebibliography}{99} % Bibliography - this is intentionally simple in this template

\bibitem{hnuc1} R. S. Hayano et al. Phys. Lett. B {\bf 231}, 355 (1989). 

\bibitem{hnuc2} T. Nagae et al. Phys. Rev. Lett. {\bf 80}, 1605 (1998).

\bibitem{hnuc3} S. Bart et al. Phys. Rev. Lett. {\bf 83}, 5238 (1999).

\bibitem{hnuc4} J. Schaffner, C. Greiner and H. St\"ocker, Phys. Rev. C
 {\bf 46}, 322 (1992).

\bibitem{hipstar1} N. K. Glendenning, Astrophys. J. {\bf 293}, 470 (1985).

\bibitem{hipstar2} J. Schaffner-Bielich, Nucl. Phys. A {\bf 804}, 309 (2008).

\bibitem{hipstar3} M. Baldo, G. F. Burgio and H. J. Schulze, Phys. Rev. C 
{ \bf 61} 055801 (2000).

\bibitem{hipstar4}  I. Vidana, A. Polls, A. Ramos, L. Engvik and M. 
Hjorth-Jensen, Phys. Rev. C {\bf 62}, 035801 (2000).

\bibitem{RH1} STAR Collab., B. I. Abelev {\it et al.},
Phys. Rev. C, {\bf 76}, (2007) 024915.

\bibitem{bu} G. Bunce {\it et al}., Phys. Rev. Lett. 
 {\bf 36}, 1113 (1976). 
 
\bibitem{hel} K. Heller {\it et al}., Phys. Lett. 
 {\bf 68B} 480 (1977); 
 K. Heller {\it et al}., Phys. Rev. Lett. {\bf 41} 607 
 (1978); 
 S. Erhan {\it et al}., Phys. Lett. {\bf 82B} 301 (1979). 

\bibitem{lu} B. Lundberg {\it et al}., Phys.Rev. D 
 {\bf 40}, 3557 (1989). 

\bibitem{wilk}  C. Wilkinson {\it et al}., Phys. Rev. 
 Lett. {\bf 46} 803 (1981). 

\bibitem{deck} L. Deck {\it et al}., Phys. Rev. D 
 {\bf 28}, 1 (1983). 

\bibitem{ram}  R. Ramerika {\it et al}., Phys. Rev. D 
 {\bf 33}, 3172 (1986). 

\bibitem{adam} M. I. Adamovich {\it et al}., Z. Phys. A 
 {\bf 350}, 379 (1995). 

\bibitem{ho} P. M. Ho {\it et al}., Phys. Rev. Lett. 
 {\bf 65}, 1713 (1990); 
 P. M. Ho {\it et al}., Phys. Rev. D {\bf 44}, 3402 (1991). 

\bibitem{mor} A. Morelos {\it et al}., Phys. Rev. Lett. 
 {\bf 71}, 2172 (1993); 
 A. Morelos {\it et al}., Phys. Rev. D {\bf 52}, 3777 
 (1995). 

\bibitem{cy} C. C. Barros Jr. and Y. Hama, Int. J. Mod. Phys. E {\bf 17}
371 (2008).

\bibitem{cy2} C. C. Barros Jr. and Y. Hama, Phys. Lett. B {\bf 699}, 
74 (2011).

\bibitem{ccb2} C. C. Barros Jr., J. Phys. Conf. Ser. {\bf 509},
012056 (2014).

\bibitem{plhc1} STAR Collab,
L. Adamczyk {\it et al.}, Nature {\bf 548}, 62 (2017).

\bibitem{BH} C. C. Barros Jr.  and Y. Hama, Phys. Rev. C  {\bf 63}, 065203 (2001).

\bibitem{cm} C. C. Barros Jr. and M. R. Robilotta, 
 Eur. Phys. J. C {\bf 45}, 445 (2006). 

\bibitem{ccb} C. C. Barros Jr., Phys. Rev. D {\bf 68}, 
 034006 (2003).

\bibitem{hyper1} A. Chakravorty {\it et al.}, Phys. Rev. 
 Lett. {\bf 91}, 031601 (2003). 

\bibitem{hyper2} M. Huang {\it et al.}, Phys. Rev. 
 Lett. {\bf 93}, 011802 (2004). 

\bibitem{manc} H. T. Coelho, T. K. Das and M. R. Robilotta, Phys. Rev. C {\bf 28}, 1812 
(1983).

\bibitem{Pil} H. Pilkuhn, {\it The Interaction of Hadrons}, (North-Holland, 
Amsterdam, 1967).

\bibitem{pi1} E. T. Osypowski, Nucl. Phys. B {\bf 21}, 615 (1970).

\bibitem{pi2}
M. G. Olsson and  E. T. Osypowski, Nucl. Phys. B {\bf 101}, 136 (1975).

\bibitem{leut1} J. Gasser, M. E. Sainio and A. \v{S}varc, Nucl. Phys. B 
{\bf 307}, 779 (1988);
T. Becher and H. Leutwyler, Eur. Phys. Journal C {\bf 9}, 643 (1999);
JHEP {\bf 106}, 17 (2001).

\bibitem{leut2} J. Gasser, H. Leutwyler and M. E. Sainio, Phys. Lett. B {\bf 253}, 252 (1991); 
{\bf 253}, 260 (1991).

\bibitem{L} A. I. L'vov, S. Scherer, B. Pasquini, C. Unkmeir and D. Drechsel, Phys. Rev. C {\bf 64}, 015203 (2001).

\bibitem{r1}  M. R. Robilotta, Phys. Rev. C {\bf 63}, 044004 (2001).

\bibitem{pdg} C. Patrignani, {\it et al.}, Chin. Phys. C {\bf 40}, 100001 (2016). 

\bibitem{stri} Stoks, V.; Rijken, T. A., Nuclear Physics A, Elsevier, v. {\bf 613}, n. 4, p. 311–341, 1997.

\bibitem{swart} Swart, J. D. Reviews of Modern Physics, APS, v. {\bf 35}, n. 4, p. 916, 1963
 
\end{thebibliography}
\end{document}